# Defect-evolved quadrupole higher-order topological nanolasers

Shengqun Guo[1], Wendi Huang[1], Feng Tian[1], Yufei Zhou[1], Yilan Wang[1], and Taojie Zhou[1,*]

[1] School of Microelectronics, South China University of Technology, Guangzhou, 510641, China
* Corresponding author: Taojie Zhou (taojiezhou@scut.edu.cn)

## Abstract

Topological photonics have been garnering widespread interest in engineering the flow of light with topological ideas. Strikingly, the recent introduction of higher-order topological insulators has generalized the fundamental framework of topological photonics, endowing counterintuitive strong confinement of light at lower-dimensional boundaries, thus unlocking exciting prospects for the exploration of topological phenomena in fresh routes as well as the design of topology-driven nanoscale light sources. Here, we revealed the photonic quadrupole topological phases can be activated by defect evolution and performed experimental demonstrations of associated nanoscale lasing operation under this paradigm. The quadrupole higher-order topological nanocavity is constructed by two topologically distinct photonic crystal slabs with opposite directions of defect evolution. Stable single mode emission and low lasing threshold in telecom C-band are achieved at room temperature of the defect-evolved quadrupole topological nanolaser. This work reveals new possibilities for photonic quadrupole topological phase transition, providing an intriguing route toward light confinement and modulation under the topological framework.

## Introduction

The past few decades have witnessed the accelerated development of the fascinating area of topological insulator (TI). In general, the representative features of TI are the topological properties of the momentum space and the robust boundary states which reflected by the bulk-boundary correspondence (*1-3*). Benefiting from the similarities between photonic crystal (PhC) and solid-state physics, analogous concepts extend into photonic systems, giving rise to the discipline of topological photonics (*4-6*). Since the demonstration of quantum Hall effect edge states (*7*), topological photonics has emerged as a remarkable platform for exploring diverse exotic topological phenomena, such as Floquet photonic TIs (*8*), quantum spin Hall photonic TIs (*9,10*), valley Hall photonic TIs (*11-13*), and phase change material induced Chern topological phase transition (*14*). In addition to the intrinsic importance for fundamental physics of topological aspect, topological photonics also provides unprecedented avenues for manipulating and confining light, thus opening a new paradigm for the design of advanced photonic devices particularly in topological lasers, including but not limited to quantum Hall lasers (*15,16*), spin-momentum-locked edge state laser (*17*), topological valley hall laser (*18*), and the topological laser based on the bulk states of quantum spin Hall insulators (*19*). Compared to the conventional laser architecture, topological lasers are prospected as a promising candidate for future on-chip high-performance coherent light sources with inherent robustness against imperfections and disorder (*20*).

Recently, a novel class of topological insulator, termed higher-order topological insulator (HOTI), has attracted significant attention due to their ability to host lower-dimensional boundary states (e.g., corner state) that go beyond the conventional bulk-boundary correspondence (*21,22*). This line of research began with the prediction of multipole TIs (*23,24*), characterized by their quantized multipole moments. To obtain quadrupole TIs in two-dimensional (2D) systems that enable the emergence of highly localized 0D corner states, the critical approach is using the appropriate distribution of positive and negative couplings to introduce π-flux (*23,24*). However, the practical implement of this approach in photonic systems remains challenging and has so far been reported in limited platforms such as coupled ring resonators (*25*) and waveguide arrays (*26*). Consequently, significant research efforts have been devoted to alternative schemes in photonic systems for the quadrupole topology beyond the π-flux mechanism. Until yet, the potential for photonic quadrupole topological phases beyond the π-flux mechanism has not been fully explored, only a few approaches have been proposed, including Floquet PhCs (*27*), magneto-optical PhCs (*28,29*), and twisted PhCs (*30*). Additionally, the promising characterizations of topological corner state provide innovative route to achieve lasing operation in an ultimate small scale with diffraction-limited mode volume ($V_m$) and ultra-low energy consumption. Nevertheless, the most extensively investigated higher-order topological corner state nanolasers are primarily based on Wannier-type HOTI with the expanded-shrunken scheme (*31-35*). The quadrupole topological nanolaser has only been demonstrated in a twisted PhC platform at a cryogenic temperature, mainly hindered by the implementation scheme of the photonics quadrupole topological phase (*36*). Naturally, we wonder if there are new possibilities for realizing the photonic quadrupole topological phase and whether it can be applied to high-performance topological nanolasers, which is fundamentally crucial for exploring intriguing topological physics within the nanoscale platforms.

In this work, we propose a defect evolution method for realizing photonic quadrupole topological phases and experimentally demonstrate its topological features within 2D photonic crystal nanolasers.

Specifically, the different branches of quadrupole topology are obtained by exploiting the evolution of the introduced geometrical defects in opposite directions (clockwise and anticlockwise). The corresponding 0D corner state is predictably located at the corner of two topologically distinguished semiconductor PhC slabs in quadrupole topology. Using InGaAsP multi-quantum wells as gain materials, we further experimentally demonstrate its stable single mode lasing operation in quadrupole higher-order topological nanolasers in telecommunication band. This work unveils a crucial method for realizing photonic quadrupole topological phase transition, establishing a new paradigm for the higher-order topological nanolaser design.

**Results**

**Defect-evolved photonic quadrupole topological phase**

Fig. 1A shows a schematic diagram of the considered defect evolution for the square lattice PhC, where the original PhC structure without introducing defects is commonly employed to describe the 2D Su-Schrieffer-Heeger model with the topological feature of Wannier-type HOTIs (*31*). Here, we consider air-hole type PhC thin slab for lasing purposes. In this scenario, the original PhC unit cell with lattice parameter *a* consists of four square-shaped air holes with length $s = 0.3a$ located at ($\pm a/4$, $\pm a/4$). Afterward, to obtain the photonic quadrupole topological phases, the geometrical defects at the diagonal position for each original air holes are introduced. The configuration describing the initial state of this process is illustrated in the left panel of Fig. 1A, where the side lengths $d_1$ and $d_2$ of introduced defects at the diagonal positions are correlated through a fixed total length $D = d_1 + d_2 = 0.5s$ and equal to $d_1 = d_2 = D/2$ in the initial state. A defect-related parameter $\Delta d_1$ (in unit of *s*) is used to describe the change of the length $d_1$ relative to the initial state in the evolved configuration. Unlike the manipulative strategy of breathing-type PhC lattices (*31,37,38*), the topological properties of this photonic system we proposed here are manipulated by the evolution of defects in different directions. When the commutative mirror symmetries $M_x$ and $M_y$ are removed by evolving the defect size either counterclockwise or clockwise, the system maintains glide symmetries $G_x = (x, y) \rightarrow (-x+a/2, y+a/2)$ and $G_y = (x, y) \rightarrow (x+a/2, -y+a/2)$, the inversion symmetry $I = (x, y) \rightarrow (-x, -y)$, and the $C_4$ symmetry, providing the possibility of the quadrupole topological bandgap.

To explore quadrupole topological phases, we present the typical transverse-electric (TE) like band structure of PhC with representative clockwise evolution of defects at $\Delta d_1 = 0.25$ as shown in Fig. 1B. For the identical magnitude of $\Delta d_1$, the clockwise and counterclockwise configurations share same band structure due to the fact that they belong to mirror-symmetric partners. Nonetheless, the bandgap for unit cell with defects evolving in different directions possesses different topological properties. To elucidate this, Wannier band polarizations are utilized to describe their quadrupole topological phases. The $j$-th Wannier bands $v_y^j(k_x)$ can be obtained from the eigenvalues $e^{2\pi i v_y^j(k_x)}$ of Wilson-loop operator $W_{k,y} = \prod_{i=0}^{n_k-1} F_{k+i\Delta k_y}$ (*24,39*), where the subscript $y$ indicates the considered path of the Wilson loop, and the matrix element for $F_k$ is $F_\kappa^{nm} = \langle u_{n,k} | u_{m,k+\Delta k_y} \rangle$, with $u_{n,k}$ being the periodic part of the Bloch wavefunction. Further, the polarizations of Wannier band can be determined by iteratively repeating the Wilson loop, i.e., nested Wilson loop along the orthogonal direction (*24*), where $u_{n,k}$ in Wilson loop is replaced by $w_{k,y}^{j} = \sum_{n=1}^{N_{occ}} u_{n,k} [\xi_{k,y}^{j}]^{n}$, which is constructed by wavefunction and *n*-th element $[\xi_{k,y}^{j}]^{n}$ of the *j*-th Wilson-loop eigenvector. Fig. 1C shows the Wannier bands for clockwise configuration with $\Delta d_1 = 0.25$. There are four gapped wannier bands labeled 1 to 4 that come in pairs above and below $v_y = 0$, indicating the vanishing total dipole moment and providing a vital piece to the emergence of the quadrupole topological phase. Noteworthy, the rich Wannier bands here can yield different combinations. The combinations "1 + 3" and "2 + 4" are presented in Fig. 1C, exhibiting the gapped composite Wannier bands. Fig. 1D shows the $k_y$-resolved polarizations $p_x^{v_y}(k_y)$ of Wannier bands. For Wannier sectors "1 + 3" and "2 + 4", they exhibit nontrivial polarization of $\pm 1/2$. Moreover, the equivalent polarization in orthogonal direction $p_y^{v_x} = p_x^{v_y}$ can be deduced via $C_4$ symmetry, thus the quantized quadrupole moment is $q_{xy} = 2 p_y^{v_x} p_x^{v_y} = 1/2$. Such quadrupole topology revealed by gapped composite Wannier bands is so-called anomalous quadrupole topology to differentiate it from conventional quadrupole topology (*30,40,41*). Meanwhile, the nontrivial quadrupole topological phases also occur in counterclockwise configuration, but they are topologically distinguished to clockwise cases by Wannier band polarization (Fig. 1E, the positive and negative signs correspond to clockwise and counterclockwise directions of evolution, respectively). These results indicate that evolution of defects in different directions in our scheme hosts distinct branches of quadrupole topology.

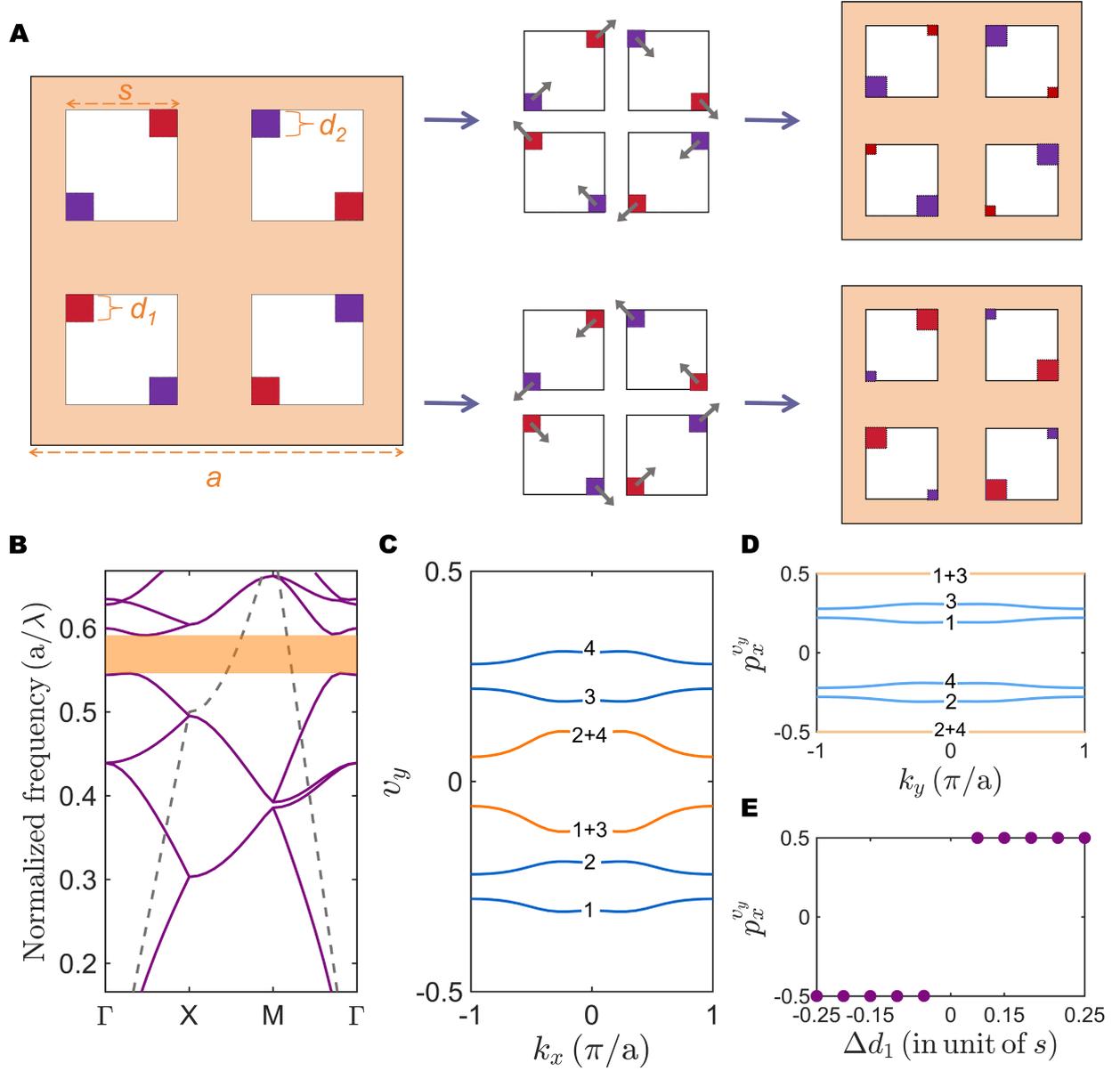

**Fig.1. Concept of the photonic quadrupole topological phase under defect evolution.** (**A**) Illustration of the defect evolution path in PhC unit cell. The gray line and yellow region represent the light line and bandgap, respectively. (**B**) Calculated TE-like photonic band structure for $\Delta d_1 = 0.25$, $a = 855$ nm, and $s = 0.3a$. (**C**) Wannier bands (blue lines) and their combinations (orange lines) for the quadrupole band gap at $\Delta d_1 = 0.25$. (**D**) The nested Wannier bands for $\Delta d_1 = 0.25$. (**E**) Evolution of nested Wannier bands under the defect-related parameter $\Delta d_1$.

**Experimental demonstration of defect-evolved quadrupole topological nanolasers**

In photonic HOTIs, a finite-size photonic system constructed by two topologically distinguished PhCs is known to support the extremely localized 0D topological corner state. To investigate the feasibility of corner state in the proposed architecture for lasing emission, the defect-evolved clockwise and counterclockwise configuration PhCs are combined to form a quadrupole topological corner state nanocavity based on previous analyses. Figure 2A shows a schematic of the defect-evolved quadrupole topological nanolaser, in which the different topological regions are identified by green and purple, respectively. The device is fabricated using the 260 nm-thick PhC slab with InGaAsP multi-quantum wells as gain medium (*42*). The scanning electron microscope (SEM) image of the fabricated device with $\Delta d_1 = 0.25$ is shown in Fig. 2B, in which the left-bottom green region corresponds to the clockwise configuration, while the remaining region is composed of the counterclockwise counterpart that evolves in the opposite direction. Figure 2C further shows a zoomed-in SEM image around the central corner boundary region.

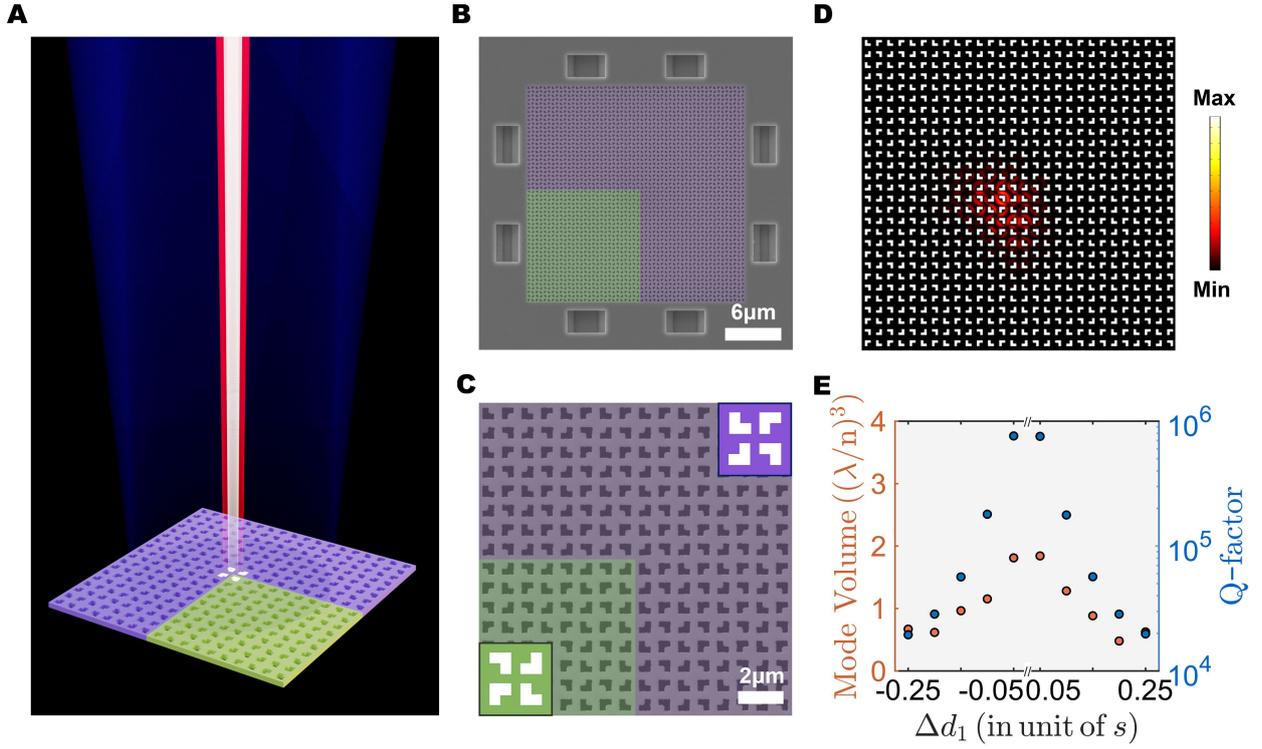

**Fig.2. Quadrupole topological nanolaser cavity design.** (**A**) Conceptual illustration of proposed defect-evolved quadrupole topological corner state nanolaser. (**B**) The top-view SEM image of the fabricated nanolaser with $a$ = 855 nm, $s$ = 0.3$a$, and $\Delta d_1$ = 0.25. (**C**) The zoomed-in SEM image around the central corner region. (**D**) The simulated electric field profiles ($|E|^2$) of the topological corner state with $\Delta d_1$ = 0.25. (**E**) Simulated $V_m$ and $Q$-factor of the topological corner state with different $\Delta d_1$.

The simulated electric field profile of the quadrupole topological corner state for a typical magnitude of $\Delta d_1$ = 0.25 is presented in Fig. 2D. As expected, the defects-evolved higher-order corner state is tightly localized around the corner region assembled by two topologically distinguished slabs. Additionally, this extreme case of defect evolution ($\Delta d_1$ = 0.25) is similar to quadrupole topological structure in acoustic systems with complex arch-shaped geometry transformation (*40,41*). To further analyze the confined corner state, the parametric dependence of the calculated $V_m$ and quality factors ($Q$-factors) on the defect-related parameter $\Delta d_1$ is shown in Fig. 2E, with $\Delta d_1$ here defined from left-bottom region, while the rest changes in the opposite direction by the same magnitude. The $Q$-factors is on the order of $10^4$ and above, which is high enough for lasing applications, and as the magnitude of $\Delta d_1$ decreases, it gradually shows the trend of the increasing $Q$-factors at the expense of sacrificing the $V_m$.

The lasing emission of the fabricated devices was characterized by the microphotoluminescence ($\mu$-PL) system at room temperature. Figure 3A presents the power-dependent measured spectra of a quadrupole topological PhC nanolaser with structural parameters $a$ = 855 nm, $s$ = 0.3$a$, and $\Delta d_1$ = 0.25. As illustrated in the figure, a sharp lasing peak (~ 1567 nm) springing from the background of spontaneous emission, and the intensity ascends as the input power increases, exhibiting the stable single mode operation over a broad range of input powers. To further demonstratethe signatures of the lasing operation, the corresponding light in-light out (*L-L*) curve and the linewidth evolution of the device are illustrated in Fig. 3B, both an obvious kink in the *L-L* curve and the linewidth narrowing effect are observed, presenting clear evidences for the stimulated emission in the fabricated quadrupole topological nanolaser. A relatively low lasing threshold around 0.5 µW is estimated form the *L-L* curve, and the linewidth near the threshold of about 0.68 nm corresponds to the experimental *Q*-factor of 2300. In contrast to the simulated *Q*-factor, the lower experimental *Q*-factor may be attributed to the unavoidable fabrication induced imperfections. For the designed quadrupole topological nanolaser, the stimulated emission from the corner state can be further verified by collecting near-field optical profiles using an InGaAs camera. Figure 3C-D depict the near-field optical profiles measured below and above lasing thresholds, respectively. The optical mode is spatially localized in the central corner region, and strong speckle pattern occurs due to the coherent emission above the threshold. Furthermore, the high-temperature stability of the proposed defect-evolved quadrupole topological nanolaser is evaluated. Fig. 3E shows the normalized measured lasing spectra under various temperatures from 30 °C to 70 °C, presenting stable single mode lasing emission as temperature increases. The collected *L-L* curves of the lasing peak under various temperatures are depicted in Fig. 3F. Figure 3G presents the experimentally determined dependence of lasing threshold on operating temperature. The lasing threshold increases rapidly under elevated temperature mainly due to the higher non-radiative recombination rates at high temperatures (*43*). These results suggest the potential practical applications for the proposed quadrupole topological nanolaser in relatively high temperature scenes.

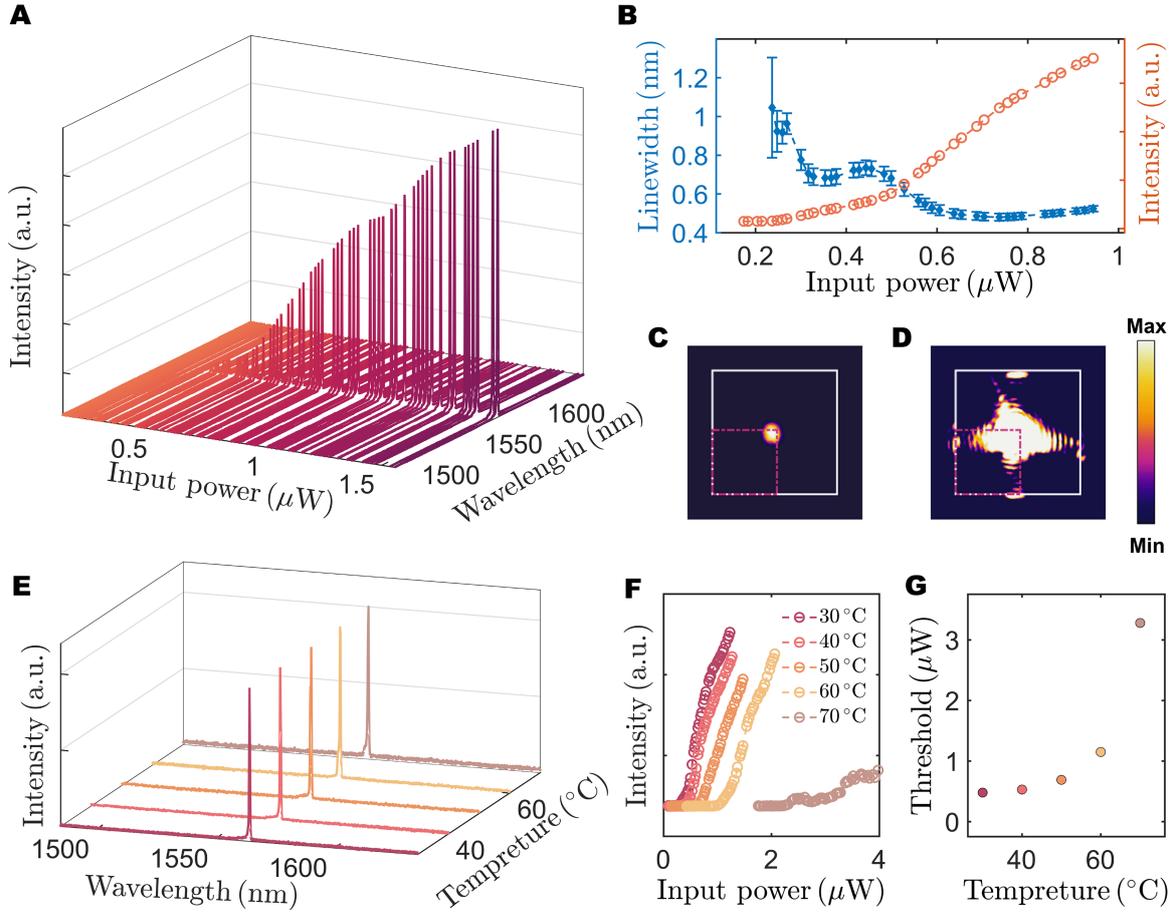

**Fig.3. Optical characterizations of the quadrupole topological nanolasers.** (**A**) Measured power-dependent emission spectra of a quadrupole topological nanolaser. (**B**) Collected *L-L* curve (orange dots) and linewidth (blue dots) under various input powers. (**C**)-(**D**) The near-field optical profiles measured below (**C**) and above (**D**) the lasing threshold. (**E**) Temperature-dependent normalized lasing spectra ranging from 30°C to 70°C. (**F**) Temperature-dependent *L-L* curves. (**G**) Lasing threshold distribution at different temperatures.

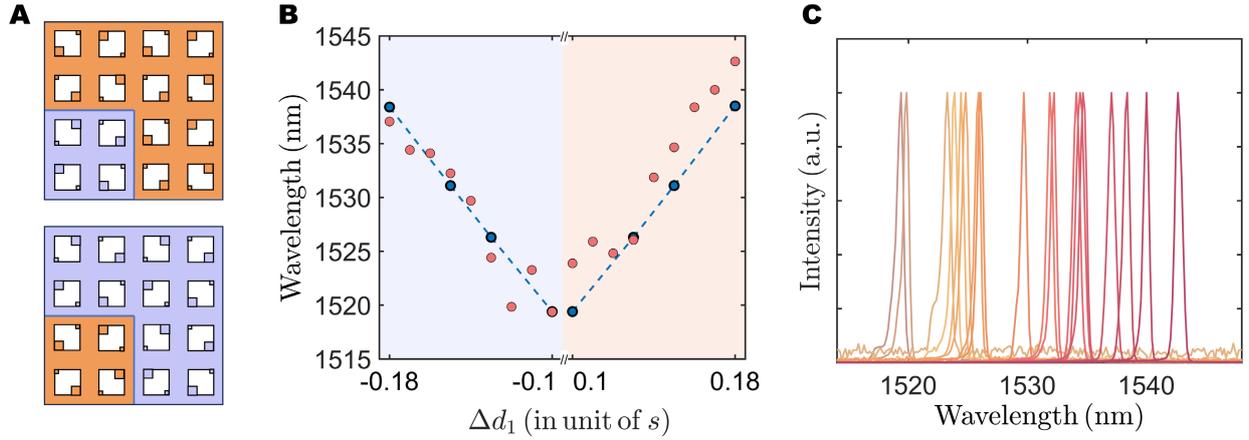

**Fig.4. Tunable wavelength manipulation under defect evolution.** (**A**) Schematic representation of defect evolution. (**B**) Simulated resonant wavelength and experimental lasing wavelength of the fabricated quadrupole topological nanolasers by varying $\Delta d_1$. The configuration of the minus/plus sign of $\Delta d_1$ corresponds to the top/bottom panel in Fig. 4A. (**C**) Normalized measured lasing spectra from nanolasers under defect evolution.

In addition to the realization of photonics quadrupole phase transition, the dynamic defect evolution also provides an opportunity for the topological nanolasers with spectrally tunable emission. Distinct from the previous prevalent wavelength manipulation strategy (*34,44*), the tunable wavelength emission of quadrupole topological nanolasers here is realized only through manipulating the defect-related parameter $\Delta d_1$ (Fig. 4A). To obtain a quantitative insight into the modulation of the output lasing wavelength upon the defect evolution, the variation of simulated wavelength (blue dots) of the corner state with different $\Delta d_1$ is shown in Fig. 4B (geometric parameter $s = 0.287a$ here to better match the experimental lasing wavelengths). It can be seen that the wavelength of the corner state increases with the magnitude of $\Delta d_1$. For the fabricated nanolasers, the lasing wavelengths with the defect-related parameter $\Delta d_1$ are also presented in Fig. 4B (orange dots). Despite irregularities and roughness of the devices induced by fabrication fluctuations, the lasing wavelengths exhibit a systematic redshift that strongly correlates with increasing $\Delta d_1$, closely aligning with the simulation trend. Figure 4C further shows the corresponding normalized lasing spectra, exhibiting that the lasing wavelengths are approximately tuned from 1519 to 1543 nm, which offers a tunable wavelength range of 23 nm. These results reveal that the geometrical defect evolution can be employed as an efficient method to achieve both precise lasing wavelength tuning of corner state and quadrupole topological phase. This approach establishes a versatile design strategy for wavelength-tunable topological nanolaser, particularly advantageous for high-throughput densely integrated photonic systems.

**Discussion**

Our study demonstrates a scheme for realizing the photonic higher-order corner state in quadrupole topological phase with introduced defect evolution as a new degree of freedom and validates them in semiconductor nanolaser. The proposed novel quadrupole topological nanolaser demonstrates stable single mode operation in the telecommunication C-band at room temperature, achieving an ultralow threshold of 0.5 μW and an experimental $Q$-factor of 2300, while maintaining robust performance up to 70°C. The corresponding tunable lasing emission is further verified by manipulating the evolution of geometrical defects. These quadrupole topological nanolasers presented here open novel prospects for the topological on-chip nanoscale coherent light sources. Meanwhile, the corner state in the quadrupole topological phase by strategically modulating the defect evolution also provides inspiration for exploring the fundamental physics and practical applications with intriguing topological properties, many possible extensions could be envisioned. For example, such topological corner state with defect evolution may find novel applications in wavelength-division multiplexing (*45*) and reconfigurable imaging (*46,47*), and it may also be able to combine with the quantum emitter for the purpose of single photon generation in topology-driven nanophotonic devices (*48*). Moreover, our results might bring about new opportunities to investigate topological rainbow trapping (*42,49,50*), large-area topological corner states (*51*), and exploring the influence of defect evolution under different topological phases may offer a possible route toward novel topological states. While the photonics quadrupole topological phase demonstrated here is based on a specific photonic system, we envisage that the proposed scheme could be generalized to other platforms, such as microwaves (*52,53*) and thermal systems (*54*), to further explore fascinating topological phenomena.

## Materials and Methods

### Device fabrication

The 260-nm-thick InP slab with six strained InGaAsP multi-quantum wells as the gain medium was used to fabricate the defect-evolved quadrupole topological nanolasers. First, plasma-enhanced chemical vapor deposition was employed to deposit 90-nm-thick $SiO_2$ as hard mask. Then, the designed topological PhC structure was defined using electron beam lithography in photoresist of PMMA950 A4 and transferred to the $SiO_2$ hard mask and further into the active layer by plasma dry etching. Afterwards, the residual resist was removed using $O_2$ plasma, and the buffered oxide etching solution was used to remove the remaining $SiO_2$ hard mask. Finally, the InP sacrificial layer is selectively removed by diluted hydrochloric acid solution, resulting in the suspended defect-evolved quadrupole topological nanolasers.

### Optical measurement

The defect-evolved quadrupole topological nanolasers were optically pumped by a 632 nm pulsed laser (pulse duration 10 ns, repetition rate 200 kHz) in μ-PL experiments. The pumping laser beam was focused on the sample using a 100× objective lens and spatially controlled by piezoelectric nanopositioners. The emission from the fabricated nanolaser was collected by the same objective lens and characterized by the spectrometer with an infrared InGaAs detector cooled by liquid nitrogen. The near-field optical images were captured by using an InGaAs camera.

### Numerical simulations

In numerical studies, the photonic band structure, $Q$-factor, $V_m$, resonant wavelength, and field profile are obtained by finite-element method. The mode volume $V_m$ of the topological nanocavity is calculated by $V_m = \frac{\int \varepsilon(\mathbf{r})|E(\mathbf{r})|^2 dV}{max(\varepsilon(\mathbf{r})|E(\mathbf{r})|^2)}$, where $\varepsilon(\mathbf{r})$ and $E(\mathbf{r})$ denote the spatially dependent dielectric constant and electric field intensity, respectively.

**Acknowledgements:** T.Z. acknowledges the startup funds from South China University of Technology. **Funding:** This work was supported by the National Natural Science Foundation of China (Grant No. 62304080), the Guangdong Basic and Applied Basic Research Foundation (Grant No. 2024A1515010802), the Science and Technology Projects in Guangzhou (Grant No. 2024A04J3683), the Fundamental Research Funds for the Central Universities (Grant No. 2023ZYGXZR068). **Author contributions:** The concepts were developed by S.G. and T.Z. S.G. performed the numerical calculations and device simulations. W.H. and F.T. fabricated the devices. S.G. and Y.Z. carried out the optical measurements with assistant from Y.W. S.G. wrote the paper with input from all coauthors. T.Z. supervised the project. **Competing interests:** The authors declare no competing interests. **Data and materials availability:** All data needed to evaluate the conclusions in the paper are present in the paper and/or the Supplementary Materials.